\documentclass[12pt,preprint]{aastex}

\slugcomment{Accepted for publication in the Astronomical Journal}

\shorttitle{Setting {\it UBVRI} Zero-Points from SDSS {\i ugriz} Magnitudes}
\shortauthors{Chonis \& Gaskell}

\begin{document}


\title{Setting {\it UBVRI} Photometric Zero-Points Using Sloan Digital Sky Survey {\it ugriz} Magnitudes}

\author{Taylor S. Chonis \& C. Martin Gaskell\altaffilmark{1}}
\affil{Department of Physics \& Astronomy, University of Nebraska,\\
Lincoln, NE 68588-0111} \email{tschonis@bigred.unl.edu,
gaskell@astro.as.utexas.edu}

\altaffiltext{1}{Present Address: Department of Astronomy,
University of Texas, Austin, TX 78712-0259. }

\begin{abstract}

We discuss the use of Sloan Digital Sky Survey (SDSS) {\it ugriz}
point-spread function (PSF) photometry for setting the zero points
of {\it UBVRI} CCD images. From a comparison with the Landolt (1992)
standards and our own photometry we find that there is a fairly
abrupt change in {\it B, V, R, \& I} zero points around $g, r, i
\sim 14.5$, and in the $U$ zero point at $u \sim 16$.  These changes
correspond to where there is significant interpolation due to
saturation in the SDSS PSF fluxes. There also seems to be another,
much smaller systematic effect for stars with $g, r \gtrsim 19.5$.
The latter effect is consistent with a small Malmquist bias. Because
of the difficulties with PSF fluxes of brighter stars, we recommend
that comparisons of {\it ugriz} and {\it UBVRI} photometry should
only be made for unsaturated stars with $g, r$ and $i$ in the range
14.5 -- 19.5, and $u$ in the range 16 -- 19.5. We give a
prescription for setting the $UBVRI$ zero points for CCD images, and
general equations for transforming from {\it ugriz} to {\it UBVRI}.

\end{abstract}


\section{Introduction}

When CCD images of a field are taken it is necessary to determine
the photometric zero points from stars of known magnitudes. It is,
however, not unusual for there to be no stars with $UBVRI$
photometry available. Fortunately, the Sloan Digital Sky Survey
(SDSS) now provides homogenous $ugriz$ photometry for stars in a
large fraction of the northern sky out of the plane of the Milky
Way.  Technical details of the SDSS are given in \citep{york00} and
\citep{stoughton02}.  The $ugriz$ system \citep{fukugita96} is
significantly different from the widely used $UBVRI$ Johnson-Cousins
system \citep{cousins76}, so it is necessary to transform between
the two systems. A number of papers
\citep{fukugita96,smith02,karaali03,karaali05,bilir05,jordi06,rodgers06,ivezic07,davenport07,bilir07}
have considered the transformations between $ugriz$ and $UBVRI$ (see
Section 6 for a discussion of these transformations).

During the course of using SDSS $ugriz$ photometry to establish the
zero points for comparison stars for photometry of active galactic
nuclei (AGN), we noticed that the zero points were different for the
fainter stars in a field than for the brighter stars.  The
difference was in the sense that stars with $g \lesssim 14$ were
systematically brighter than predicted from the SDSS magnitudes. The
difference did not seem to depend on the color of the stars and a
check of the CCD used showed no evidence for non-linearity.  A
subsequent comparison of magnitudes of Landolt standards
\citep{landolt92} revealed a similar zero-point difference for stars
brighter or fainter than $r \sim 14$.

In this note we report results of our investigation of the
limitations of using SDSS photometry for bright stars, and give a
prescription for setting zero points in CCD images taken through
{\it UBVRI} filters.

\section{Transformation Equations}

We obtained {\it ugriz} magnitudes from SDSS data release 5\footnote{See
\it{http://www.sdss.org/dr5/}} (DR5) \citep{abazajian05} for the \cite{landolt92}
standard stars in SDSS fields. We first removed very blue and red stars
outside the ranges $0.08 < (r-i) < 0.5$ and $ 0.2 < (g-r) < 1.4$. We then
plotted the $(r-i)$ vs. $(g-r)$ color-color diagram and removed outlying points
more than 2.5 standard deviations from the linear least squares fit.
We derived transformation equations only for stars with $r > 14$. A
few points lying more than 2.5 standard deviations away from the
least-squares fits were removed. We obtained the following
transformations:

\begin{equation}
B = g + (0.327 \pm 0.047) (g-r) + (0.216 \pm .027)
\end{equation}

\begin{equation}
V = g - (0.587 \pm 0.022) (g-r) - (0.011 \pm .013)
\end{equation}

\begin{equation}
R = r - (0.272 \pm 0.092) (r-i) - (0.159 \pm .022)
\end{equation}

\begin{equation}
I = i - (0.337 \pm 0.191) (r-i) - (0.370 \pm .041)
\end{equation}

As is well known, tranformations to $U$ are particularly
problematic. Since our aim is only to give a prescription for
setting $UBVRI$ zero points rather than to obtain transformations
valid for individual stars for astrophysical purposes, we determined
the transformation for the $U$ filter as follows.  First we removed
all stars that were more than 2.5 standard deviations from a linear
fit in four-dimensional $(u-g), (g-r), (r-i), (i-z)$ color space.
For the remaining stars with no saturation warning flags, we
restricted ourselves to stars with $1 < (u-g) < 2$ and $u
> 16$. For these stars we found no statistically significant
dependence on the $(u-g)$ color.  This is not surprising since, of
the SDSS $ugriz$ filters, the passband of the $u$ filter agrees most
closely to the Johnson-Cousins passbands.  The transformation for
$U$ is thus

\begin{equation}
U = u - 0.854 \pm 0.007
\end{equation}

\section{Magnitude Dependencies}

In Figs. 1 -- 5 we show the dependencies of the differences between
the $UBVRI$ magnitudes observed by \cite{landolt92} and those
calculated using equations (1--5) versus $u$, $g$, $r$, or $i$
color\footnote{Since we have restricted ourselves to stars with a
fairly narrow range of color falling near a linear $(r-i)$ vs.
$(g-r)$ relationship, Figs. 1 -- 5 look similar if a different $ugriz$ filter
is plotted on the horizontal axis.}.  It can be seen that in each
case the SDSS magnitudes under predict the $BVRI$ magnitudes by
$\sim 0.15$ mag. for the brighter stars and the $U$ magnitude by up
to $\sim 2$ magnitudes. In Fig. 1 we also show the residuals in $B$
for stars for which we obtained $B$-band photometry as part of our
AGN monitoring program. The systematic differences we see for these
stars are consistent with those found from the \cite{landolt92}
standards. Since our CCD photometry was obtained with a completely
different setup from the \cite{landolt92} photomultiplier
photometry, the agreement removes the possibility that the magnitude
dependency is due to a hitherto undetected systematic effect in the
\cite{landolt92} photometry. The effect must arise instead from the
calculation of PSF magnitudes in the SDSS data reduction pipeline
when there is saturation of bright stars. In Figs. 1--5 we have
indicated with crosses and triangles which stars have saturation
warning flags in the SDSS data base. Clearly, any star with a
saturation warning flag associated with it should not be used for
determining photometric zero points. In addition, extreme caution
should be used when using photometry of these saturated stars in
any application. It is interesting that apart from the abrupt change
at $g, r, i \sim 14$ the PSF magnitudes are surprisingly good up
to $g, r, i \sim 11$.

\section{Faint Stars}

Although our main concern in this note has been to investigate SDSS
photometry of bright stars, we also looked for systematic effects at
faint levels.  \cite{jordi06} have derived transformations between
$griz$ and $BVRI$ photometry for a large number of stars (see Jordi
et al. for a description of the data sources).  Their data set is
inhomogeneous but includes a number of faint stars ($V > 20)$.  The
\cite{jordi06} data show a large scatter (see figures in their
paper) so in Fig. 6 we show mean $V_{Johnson} - V_{SDSS}$ residuals
as a function of $g$ for these data.  For $14.5 < g < 17.5$ the mean
residuals are, on average, close to zero (the horizontal line).
\cite{jordi06} exclude stars with $r < 14$ because of concerns with
saturation effects, but it can be seen that the average residuals
show a systematic deviation in the brightest bin ($g = 14.25)$ in
the same sense as we find in Fig. 2 (but note that the scale in Fig.
6 has been magnified by a factor of ten). There is also a systematic
deviation for $g > 19.5$. This systematic deviation is smaller,
however, than the effect we find at $g \sim 14$.

One factor in the turndown for $g > 19.5$ in Fig. 6 could be
Malmquist bias.\footnote{We are grateful to the referee for
suggesting this possibility.} Comparing the quoted standard errors
for the photometry in the data set used by \cite{jordi06} with the
errors in the $g$-band SDSS photometry shows that the SDSS
photometric standard errors are several times larger than the
standard errors in the other photometry at faint magnitudes.  We
simulated the effects of Malmquist bias with Monte Carlo simulations
by creating artificial $BVRI$ photometry of 1000 faint SDSS stars
whose magnitudes had Gaussian noise added to them that was
proportional to the quoted SDSS standard errors.  The thin solid
line in Fig. 5 shows the effect of the bias.  This should be
regarded as a lower limit to the Malmquist bias.  The slope of the
Malmquist bias will increase with increasing random differences
between the SDSS magnitudes and the $BVRI$ photometry, and could
easily be twice as great as shown.  In addition to the effects of
Malmquist bias there could be small systematic differences in the
transformations and zero points for the inhomogenous data sets used
by Jordi et al. Different populations of stars at faint magnitudes
could also be a factor. These other uncertainties could be the cause
of the very slight systematic effect over the intermediate range $15
< g < 20$.

\section{Determining Zero Points in CCD Images}

We offer the following prescription for determining the zero points
of CCD images taken through standard $UBVRI$ filters.  After
instrumental magnitudes have been determined for all stars in the
field, the stars are matched up with stars returned by the SDSS
Skyserver.\footnote{{\it
http://cas.sdss.org/astro/en/tools/search/radial.asp}}  A $(g-r)$
vs. $(r-i)$ color-color plot should be made for all stars without
saturation warning flags.  The list should then be cleaned to remove
all stars whose colors lie outside the ranges given in section 2
above, and which lie far from the linear color-color relationship.
The needed $UBVRI$ magnitudes are then found using equations (1) --
(5), and the mean photometric zero points set using the average from
the fainter stars.

\section{Discussion}

The transformations we give in equations (1) -- (5) are consistent
with the range of previously published transformations. Because our
linear transformation equations are derived for a practical purpose
of calibrating $UBVRI$ photometry, and are available for each of the
Johnson-Cousins filters individually, our transformations are
different in nature from those previously published.  We briefly
summarize here previously published transformations and discuss how
they differ from the ones given above. \cite{fukugita96} give
synthetic transformations from $UBVRI$ to $u'g'r'i'z'$.
\cite{smith02} gave transformations between $UBVRI$ and $u'g'r'i'z'$
magnitudes observed with the Photometric Telescope (PT) at Apache
Point Observatory for some filters and for colors. \cite{rodgers06}
give improved color transformations between $u'g'r'i'z'$ and $UBVRI$
for main-sequence stars. They also consider higher-order color
terms. It is important to note the difference between $u'g'r'i'z'$
and $ugriz$. This is discussed in \cite{smith07}. Additional
technical details concerning the difference between the two systems
as well as transformations between them are discussed in
\cite{tucker06}. \cite{jordi06} give color transformations between
$ugriz$ as observed with the SDSS 2.5-m telescope (rather than the
PT) and $UBVRI$. Additional transformations are given by
\cite{jester05}, \cite{karaali03}, \cite{karaali05},\cite{bilir05},
\cite{davenport07}, and \cite{bilir07}. Some of the transformations
including \cite{ivezic07} consider polynomials in the color terms,
but we found no need for higher-order terms for the restricted range
of colors we consider. Note that the above cited transformations
consider only colors, transform from $UBVRI$ to $ugriz$, are derived
for the $u'g'r'i'z'$ system, or give transformations only for select
Johnson-Cousins filters.

In this note, our aim has been to give a practical means of
photometrically calibrating $UBVRI$ CCD images. Researchers who are
interested in astrophysical applications of SDSS photometry (such as
the determination of spectroscopic parallaxes or fitting theoretical
isochrones to HR diagrams) are referred to the above mentioned
papers because the $ugriz$ to $UBVRI$ transformations depend on the
luminosity class and metallicity of the stars.  We have minimized
these effects for zero-point setting by using a fairly tight color
selection.

\acknowledgments

We are grateful to Katrin Jordi for supplying the data from Jordi et
al. (2006) in machine-readable format, to Robert Lupton for useful
discussion of the SDSS handling of PSF saturation, and the referee
for useful comments.  We wish to thank Tom Miller for making the
photometric observations possible. This research has been supported
by National Science Foundation grant through AST 03-07912, and the
University of Nebraska UCARE program.

\clearpage

\begin{figure}
\epsscale{.80} \plotone{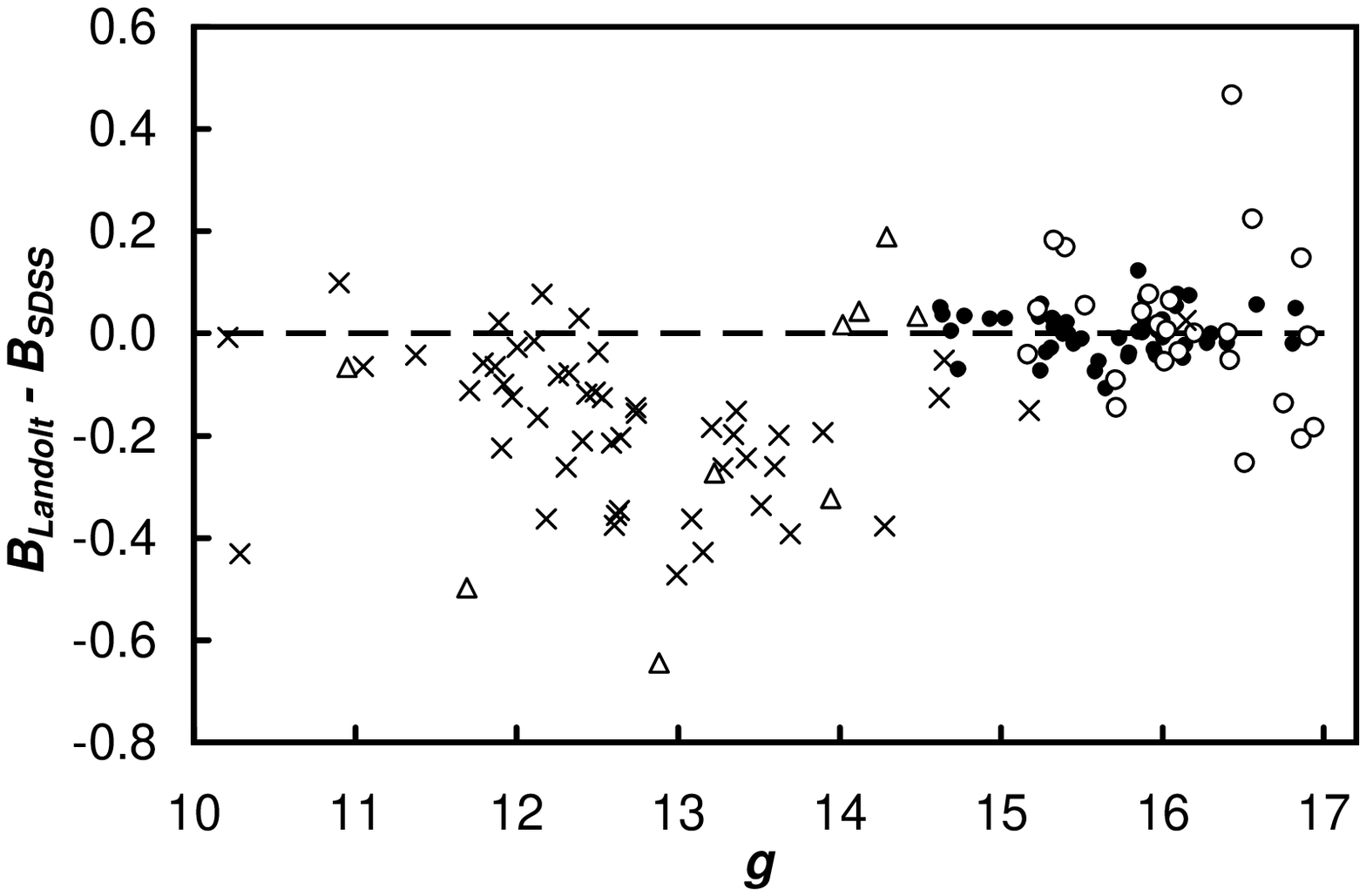} \caption{Residuals of observed
Landolt $B$ magnitudes minus $B$ predicted from SDSS $ugriz$
photometry versus $g$. The filled circles and crosses are stars from
\cite{landolt92}. Stars with SDSS saturation warning flags are shown
as crosses. The open circles and triangles are from our CCD
photometry with a 0.4-m telescope. Stars with saturation warning
flags are shown as triangles. The horizontal dashed line shows the
zero point determined from the fainter stars.\label{fig1}}
\end{figure}

\begin{figure}
\epsscale{.80} \plotone{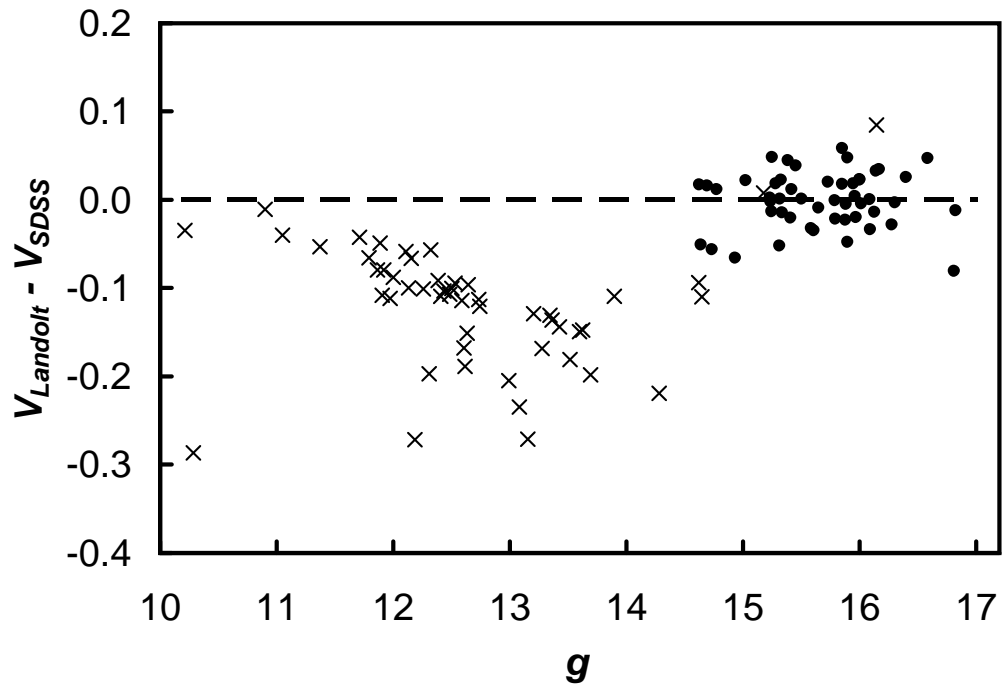} \caption{Residuals of observed
Landolt $V$ magnitudes minus $V$ predicted from SDSS $ugriz$
photometry plotted against $g$. Symbols are as in Fig. 1. The
horizontal dashed line shows the zero point determined from the
fainter stars.\label{fig2}}
\end{figure}

\begin{figure}
\epsscale{.80} \plotone{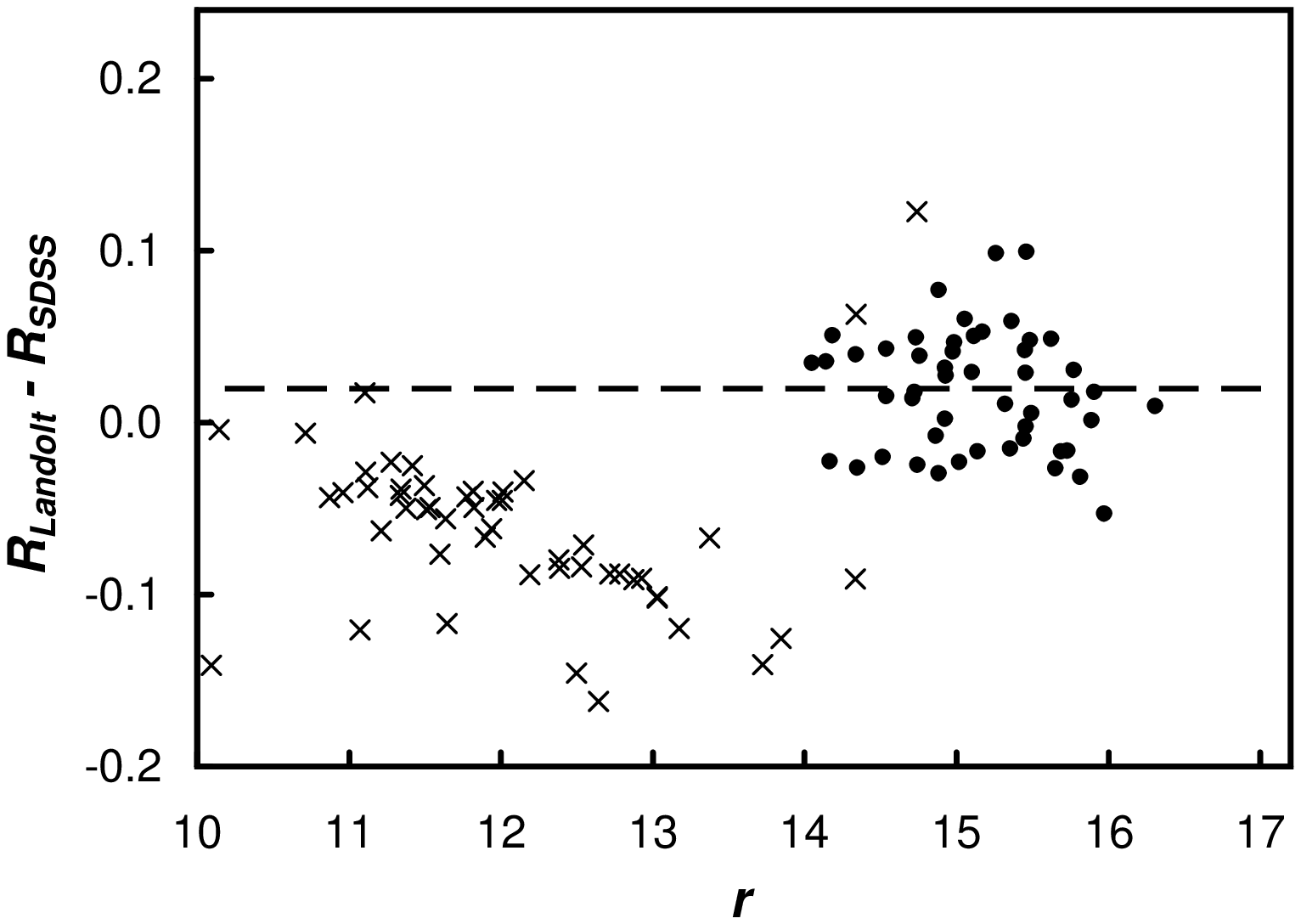} \caption{Residuals of observed
Landolt $R$ magnitudes minus $R$ magnitudes predicted from SDSS
$ugriz$ photometry plotted against $r$. Symbols are as in the
previous figures. The horizontal dashed line shows the zero point
determined from the fainter stars. \label{fig3}}
\end{figure}

\begin{figure}
\epsscale{.80} \plotone{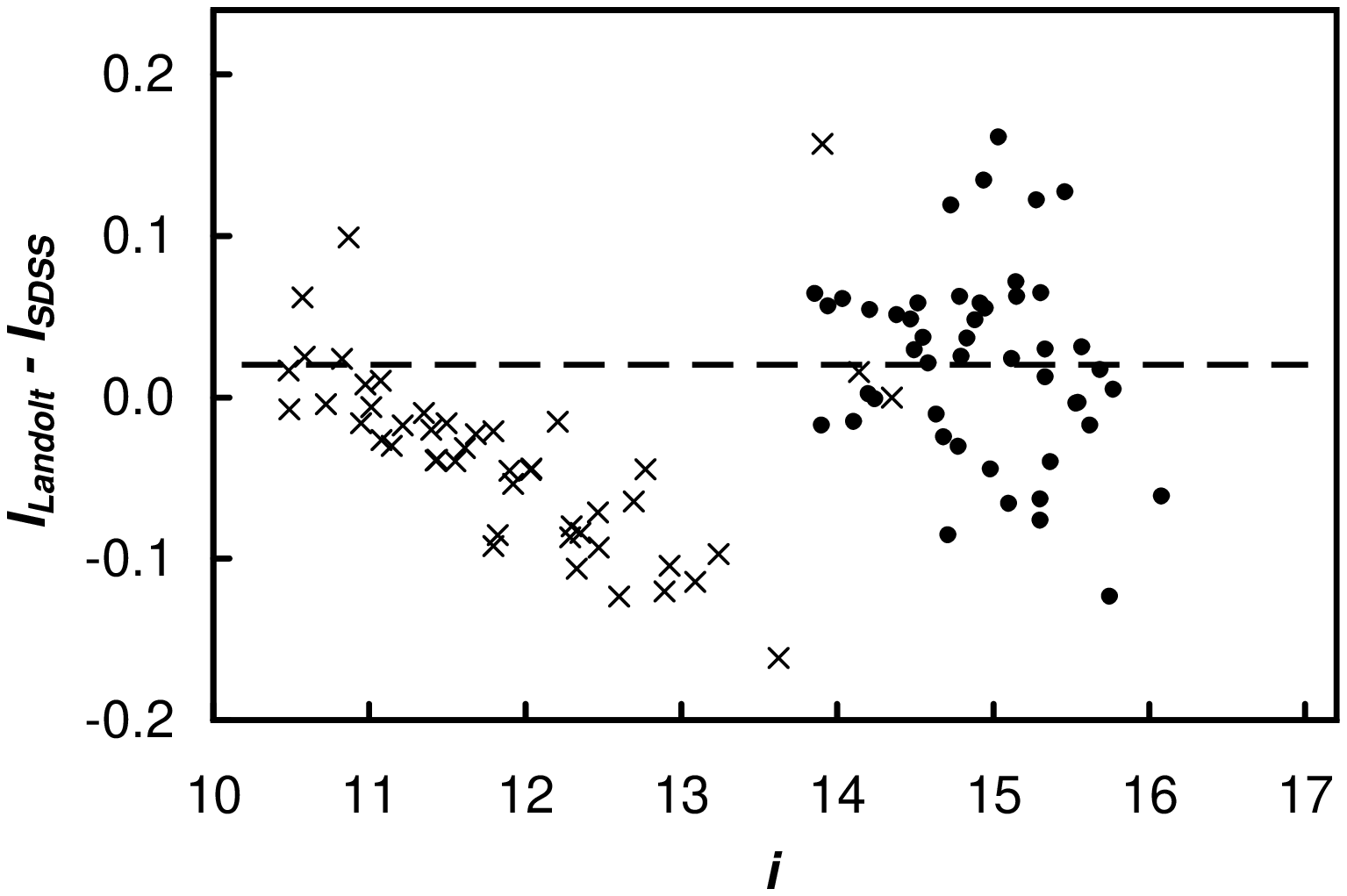} \caption{Residuals of observed
Landolt $I$ magnitudes minus $I$ magnitudes predicted from SDSS
$ugriz$ photometry versus $i$. Symbols are as in the previous
figures. The horizontal dashed line shows the zero point determined
from the fainter stars. \label{fig4}}
\end{figure}

\begin{figure}
\epsscale{.80} \plotone{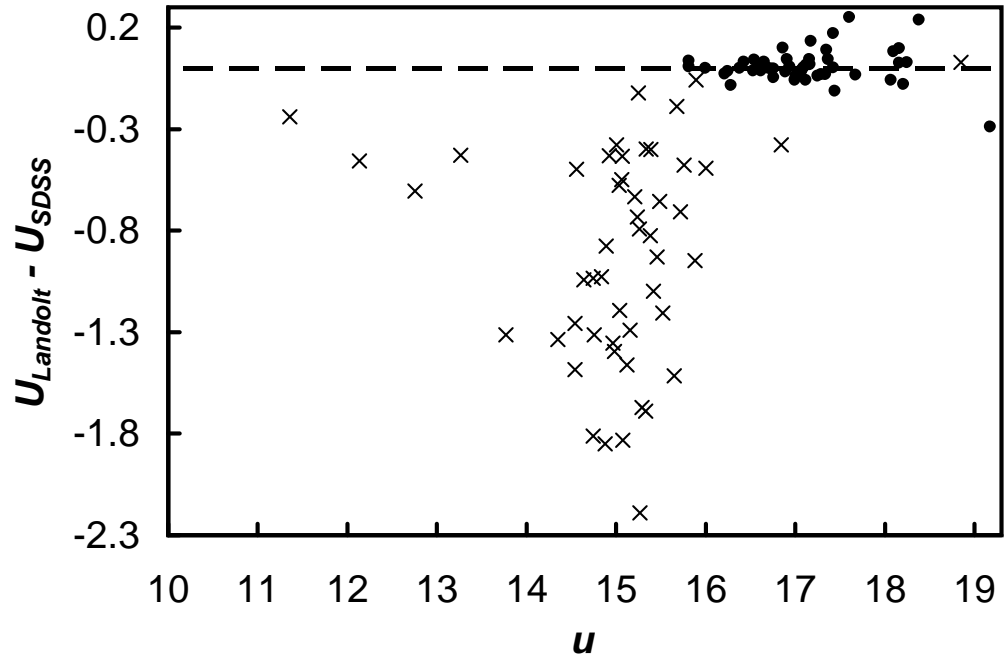} \caption{Residuals of observed
Landolt $U$ magnitudes minus $U$ predicted from SDSS $u$ photometry
versus $u$. Symbols are as in the previous figures. The horizontal
dashed line shows the zero point determined from the fainter stars.
Note that the vertical scale is larger than in the previous figures
because of the larger standard errors associated with the $u$
magnitude saturation.\label{fig5}}
\end{figure}

\begin{figure}
\epsscale{.80} \plotone{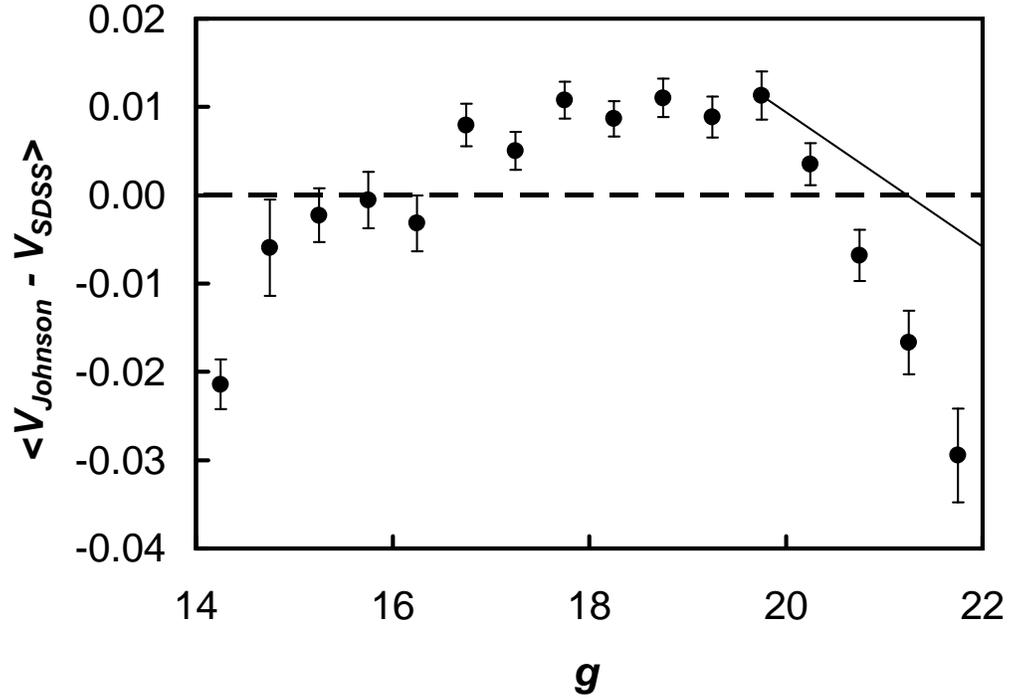} \caption{Mean residuals of observed
$V$ magnitudes minus $V$ magnitudes predicted from SDSS $ugriz$
photometry shown as a function of $g$ for the large photometric
sample compiled by \cite{jordi06}.  The error bars show the standard
errors in the means. Note that the vertical axis scale is magnified
ten times compared with the scales in Figs. 1 -- 4.  The horizontal
dashed line shows the expected relationship if there are no
magnitude-dependent effects. The thin solid line shows the expected
effect of a Malmquist bias (see text).  \label{fig6}}
\end{figure}

\end{document}